\documentclass[11pt,twoside]{article}
\usepackage{cal05}

\contact{Jason Rhodes}
\email{jason.d.rhodes@jpl.nasa.gov}

\begin{document}

%
%
%

\title{Modeling and Correcting the Time-Dependent ACS PSF}

%
%
%

\author{Jason Rhodes}
\affil{Jet Propulsion Laboratory, 4800 Oak Grove Drive, Pasadena, CA 91109}
\author{Richard Massey, Justin Albert, James E. Taylor}
\affil{California Institute of Technology, 1200 East California Blvd, Pasadena, CA 91125}
\author{Anton M. Koekemoer}
\affil{Space Telescope Science Institute, 3700 San Martin Drive, Baltimore, MD, 21218}
\author{Alexie Leauthaud}
\affil{Laboratoire d'Astrophysique de Marseille, BP 8, Traverse
du Siphon, 13376 Marseille Cedex 12, France}


%

%
%
%

\paindex{Rhodes, J.D.}
\aindex{Massey, R.}
\aindex{Albert, J.}
\aindex{Taylor, J.E.}
\aindex{Koekemoer, A. M.}
\aindex{Leauthaud, A.}

%
%

\authormark{Rhodes, Massey, Albert, Taylor, Koekemoer \& Leauthaud}



\def\multidrizzle{{\it MultiDrizzle}}
\def\tinytim{{\it TinyTim}}

\begin{abstract} The ability to accurately measure the shapes of faint objects
in images taken with the {\it Advanced Camera for Surveys} (ACS) on the {\it
Hubble Space Telescope} (HST) depends upon detailed knowledge of the Point
Spread Function (PSF). We show that thermal fluctuations cause the PSF of the ACS
Wide Field Camera (WFC) to vary over time. We describe a modified version of the
\tinytim\ PSF modeling software to create artificial grids of stars across the
ACS field of view at a range of telescope focus values. These models closely
resemble the stars in real ACS images. Using $\sim10$ bright stars in a real
image, we have been able to measure HST's apparent focus at the time of the
exposure. \tinytim\ can then be used to model the PSF at any position on the
ACS field of view. This obviates the need for images of dense stellar fields at
different focus values, or interpolation between the few observed stars. We show
that residual differences between our \tinytim\ models and real data are
likely due to the effects of Charge Transfer Efficiency (CTE) degradation.
Furthermore, we discuss stochastic noise that is added to the shape of point
sources when distortion is removed, and we present \multidrizzle\ parameters that
are optimal for weak lensing science. Specifically, we find that reducing the
\multidrizzle\ output pixel scale and choosing a Gaussian kernel significantly
stabilizes the resulting PSF after image combination, while still eliminating
cosmic rays/bad pixels, and correcting the large geometric distortion in the
ACS. We discuss future plans, which include more detailed study of the effects
of CTE degradation on object shapes and releasing our \tinytim\ models to the
astronomical community. \end{abstract}


\keywords{ACS/WFC}
\keywords{point spread function (PSF):ACS}
\keywords{MultiDrizzle}


\section{Introduction and motivation}

Accurate shape measurements of faint, small galaxies are crucial for certain
applications, most notably the measurement of weak gravitational lensing.
Quantifying the slight distortion of background galaxies by foreground matter
relies on detecting small but coherent changes in the shapes of many galaxies
(see Refregier 2003 for a recent review). To extract the lensing signal, it is
crucial to remove instrumental effects from galaxies' measured shapes. On the
{\it Hubble Space Telescope} (HST), these include:

\newpage

\begin{itemize}
\item Convolution of an image with the telescope's Point Spread Function (PSF).

\item Geometric distortion of an image. This is particularly large in the {\it
Advanced Camera for Surveys} (ACS) because of its location off HST's optical axis.

\item
Trailing of faint objects in the CCD readout direction due to degraded Charge Transfer Efficiency (CTE).
\end{itemize}

\noindent In this proceeding, we describe a method to model and correct for the
telescope's temporally and spatially varying PSF.  The geometric distortion has
already been shown to be successfully removed during image processing by \multidrizzle\ (Koekemoer et al.\ 2002). Removing the distortion does change the PSF,
and we present recommendations to minimize stochastic changes introduced during the
repixellization  stage of image processing. The effect  of continuing CTE degradation on galaxy
shapes is only becoming apparent as the ACS spends longer in orbit, and is not
yet completely understood. That is therefore beyond the scope of this
proceeding. A separate method to remove CTE effects will be presented in Rhodes
et al.\ (2006), and the application of all these corrections in a weak lensing
analysis will be presented in Massey et al.\ (2006). Other branches of
astronomy, including stellar photometry in crowded fields, the study of AGN, and
proper motions also require detailed knowledge of the PSF and will benefit from
the models we describe here.

\begin{figure}
\plottwo{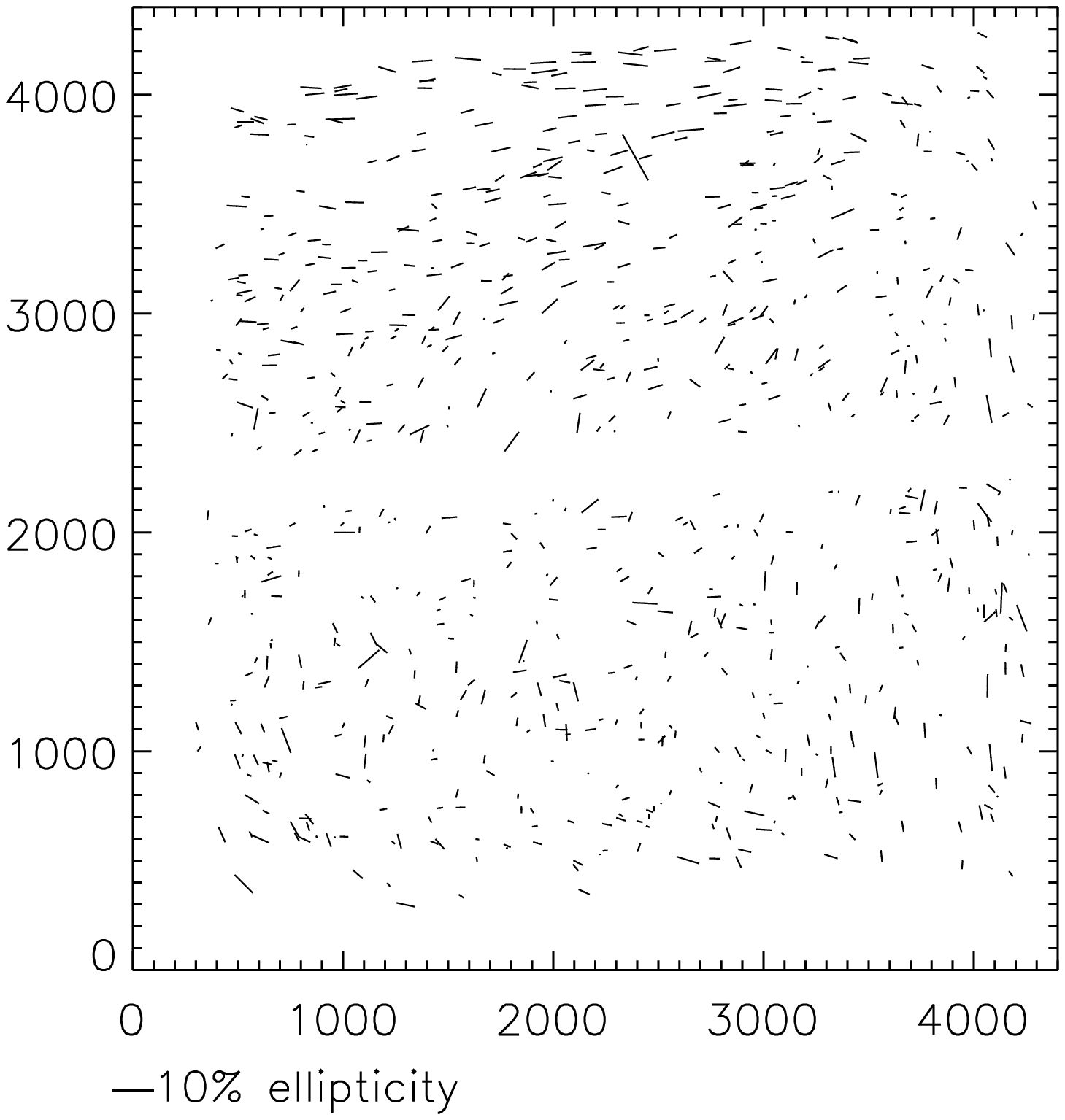}{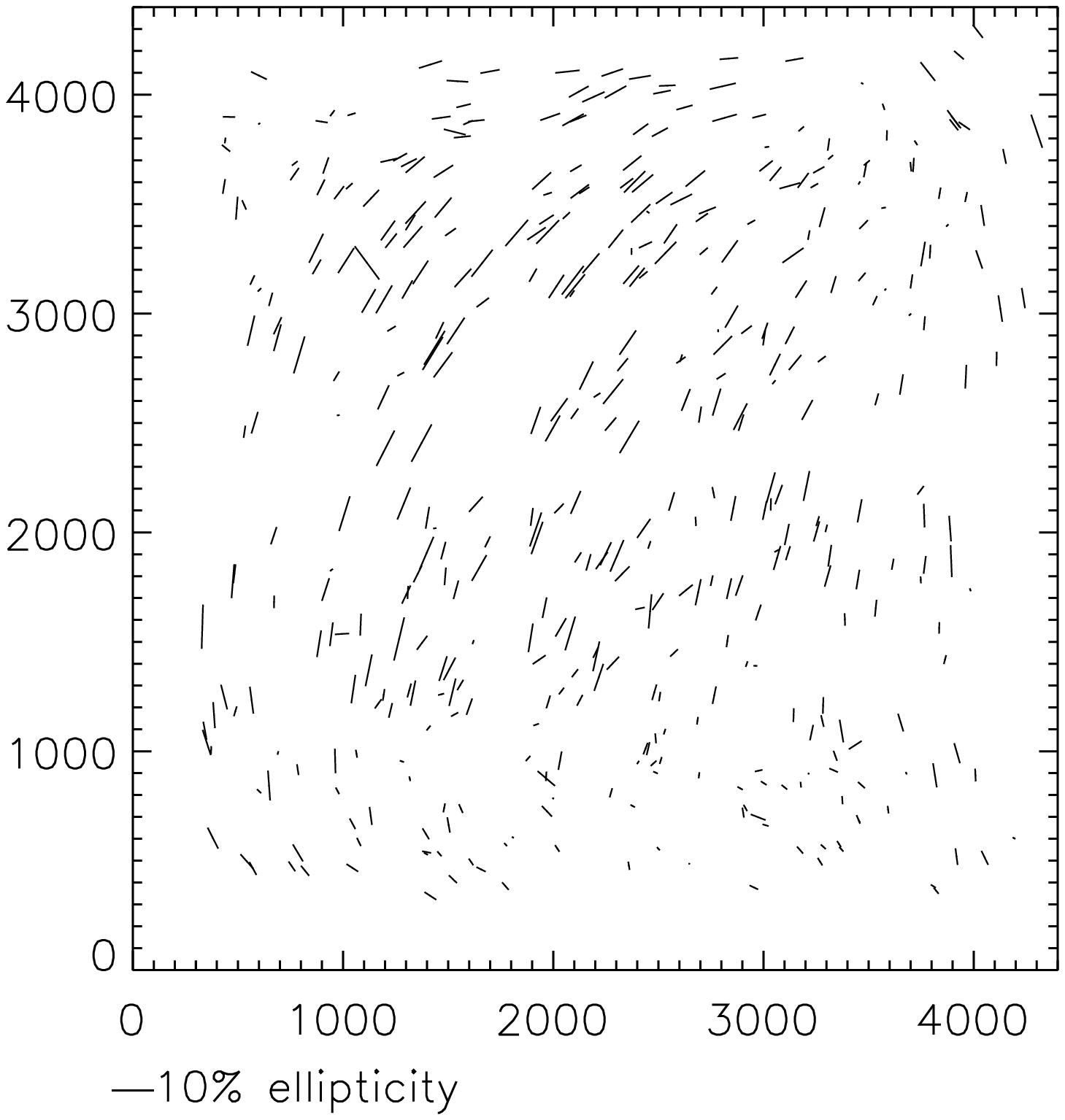}
%
%
\caption{The ellipticity of stars in the COSMOS survey observed with the ACS WFC  while HST happened to be at nominal (left panel) and low (right
panel) focus. The orientation and size of each tick mark represents the
ellipticity of one star; both panels contain stars from several different
fields. The difference in the PSF patterns is apparent, and demonstrates the
need for a time dependent PSF model.} \label{fig:cosmospsfs} \end{figure}

In weak lensing, to deconvolve galaxy shapes from the PSF, we must accurately
know the shape of the PSF at the position of the object and at the time of the
observation For example, see Rhodes, Refregier \& Groth (2000) for a description of the method
we use on the Cosmic Evolution Survey (COSMOS; Scoville et al.\ 2006) images we use to test the PSF models we describe in this paper.
 If the HST PSF were stable over time, it would be
 straightforward  to build a catalog of stellar images across
the entire field of view. However, thermal fluctuations in HST that change its
effective focus (the distance between the primary and secondary mirrors) lead to
temporal PSF variations. As an example, Figure~\ref{fig:cosmospsfs} shows the
PSF pattern in two sets of COSMOS images.
The left hand panel shows stars from images taken when the telescope was near optimal focus, and the right hand panel shows stars observed when the telescope was several microns below optimal focus.
 Each
tick mark in the figure represents the ellipticity of one star, measured using
the standard weak lensing definition,

\begin{equation}
\label{ellipticity}
|e|=\frac{\left[(I_{xx}-I_{yy})^2+(2I_{xy})^2\right]^{1/2}}{I_{xx}+I_{yy}} ~,
%
%
%
\end{equation}

\noindent where star's weighted second order moments

\begin{equation}
\label{moments}
I_{ij}=\frac{\sum w I x_i x_j }{\sum w I}
\end{equation}

\noindent involve summations over all pixels. $I$ is the intensity of a pixel,
$w$ is some weighting function (in our case a Gaussian with a width of about the
FWHM of the PSF), and $x_i$ is the distance of a pixel from the centroid of the
object. It is apparent from Figure~\ref{fig:cosmospsfs} that changes in the PSF
over time are sufficient for a temporally stable ACS PSF model to be inadequate
in demanding applications, when using data collected over a period of more than
a few days. Other effects, including CTE, introduce additional variation on
longer time scales.


This proceeding is organized as follows. In \S\ref{TinyTim} we introduce the
\tinytim\ software package that we used for PSF modeling, and discuss
modifications that we have made. In \S\ref{multidrizzle} we discuss \multidrizzle\
and how to minimize the aliasing of point sources that occurs during distortion
removal. In \S\ref{models} we show  how we have used our \tinytim\ models to quantify the temporal variation of the ACS PSF, and describe
how the same models can be used to correct for it. In \S\ref{conclusions} we
draw conclusions and outline a plan to release our \tinytim\ PSF models.

\section{TinyTim PSF models}
\label{TinyTim}

We have adapted version 6.3 of the \tinytim\ software package (Krist \& Hook
2004) to create simulated images of stars. \tinytim\ creates FITS images
containing one or more stars that include the effects of diffraction, geometric
distortion, and charge diffusion within the CCDs. By default the images appear
as they would in raw ACS data; they are highly distorted and have a pixel scale
of 0.05 arcseconds per pixel. We have written an IDL wrapper to undo the
distortion, resample the images, and combine adjacent PSFs to mimic the effects of
dithering. 
The wrapper can also run \tinytim\ multiple times and create a grid of
PSF models across the whole ACS field of view. We insert our artificial stars
into blank images with the same dimensions and FITS structure as real ACS data,
thereby manufacturing arbitrarily dense starfields.

This basic pipeline calculates a diffraction pattern (spot diagram), distorts
it, and adds charge diffusion; all three effects usually depend upon the
position of the star in the ACS field of view. We have made two versions of
artificial starfields with important changes to this basic pipeline. The
deviations from this default are:

\begin{itemize}

\item In order to examine the effects of the distortion removal process, we have
created a version of our \tinytim\ starfields where each star has an identical
diffraction pattern and charge diffusion, but a geometric distortion determined
by the location of the PSF within the ACS field of view. Once the geometric
distortion is removed (for example by running the field through \multidrizzle),
these stars should all appear identical.

\item In order to correct data, we have created a second version of our \tinytim\ starfields that do not contain the effects of geometric distortion at
all, instead modeling stars as they would appear after a perfect removal of
geometric distortion. Conversion between
distorted and non-distorted frames, which is necessary to simulate charge
diffusion in the raw CCD, was performed using very highly oversampled images.
This avoids stochastic aliasing of the PSF (see \S\ref{multidrizzle}), and
minimizes noise in the PSF models.

\end{itemize}

\noindent We describe the application of these simulations in the following two sections.

\section{Optimization of MultiDrizzle parameters for Weak Lensing Science}
\label{multidrizzle}

\multidrizzle\ is used to combine dithered exposures, remove cosmic rays and
bad pixels, and eliminate the large geometric distortion in ACS WFC images
(Koekemoer et al.\ 2002). However, the transformation of pixels from the distorted input image to the undistorted output plane can introduce significant ``aliasing''
of pixels if the output pixel scale is comparable to the input scale. When
transforming a single input image to the output plane, point sources can be
enlarged, and their ellipticities changed by several percent, depending upon
their sub-pixel position. This is one of the fundamental reasons why dithering
is recommended for observations,
since the source is at a different sub-pixel position in different exposures,
hence the effects are mitigated to some extent when all the exposures are
combined. However, the remaining effect in combined images is still quite
sufficient to prevent the measurement of small, faint galaxies at the
precision required for weak lensing analysis.

%
%

Of course, such pixellization effects are unavoidable during the initial
exposure, when  the detector discretely samples an image. However, it is clearly
desirable to minimize related effects during data reduction. The effect on each
individual object depends on how the input and output pixel grids line up.
Indeed, this can be mitigated by using a finer grid of output pixels ({\it
e.g.} Lombardi et al. 2005). The reduction in pixel scale (which will cause a
corresponding increase in computer overheads) can be performed in conjunction
with simultaneously ``shrinking'' the area of the input pixels that contains
the signal, by making use of the \multidrizzle\ \texttt{pixfrac} parameter
and convolution kernel.

\begin{figure}
\epsscale{0.5}
\plotone{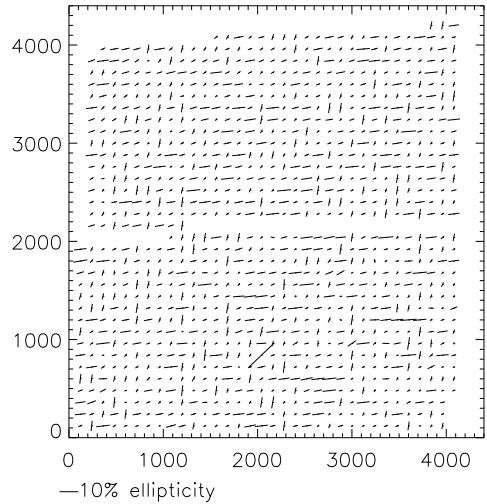}
\caption{``Aliasing'' of the PSF introduced when transforming a single distorted
input image to the undistorted output frame. The tick marks represent the ellipticity of stars that have undergone identical diffraction in
the telescope's optics and should therefore look identical. The only difference
between stars is their sub-pixel position when their geometric distortion is
removed and the images are combined. The apparent difference between the tick
marks shows that the PSF is changed. The problem can be ameliorated by altering
several of the \multidrizzle\ settings, in particular reducing the output pixel
scale.}
\label{fig:aliasing}
\end{figure}

We have run a series of tests on the simulated PSF grids described in
\S\ref{TinyTim} to determine the optimal values of the \multidrizzle\
parameters specifically for weak lensing science. To this end, we first produced a grid of stars that ought to look
identical after the removal of geometric distortion. Figure~\ref{fig:aliasing}
shows the ``aliasing'' produced when the distortion is removed from a single
image. We have also created a series of four dithered input images (with
the linear dither pattern used for the COSMOS survey); the scatter in the
ellipticities of the output stars is then approximately half as big. This
confirms the idea that the repixellization adds stochastic noise to the
ellipticity when the four sub-pixel positions are uncorrelated. For weak
lensing purposes, this noise is still substantial. With enough dithered input
images, the scatter of ellipticities could be reduced further, but this is not
feasible for most observations.

\begin{figure}
\plotone{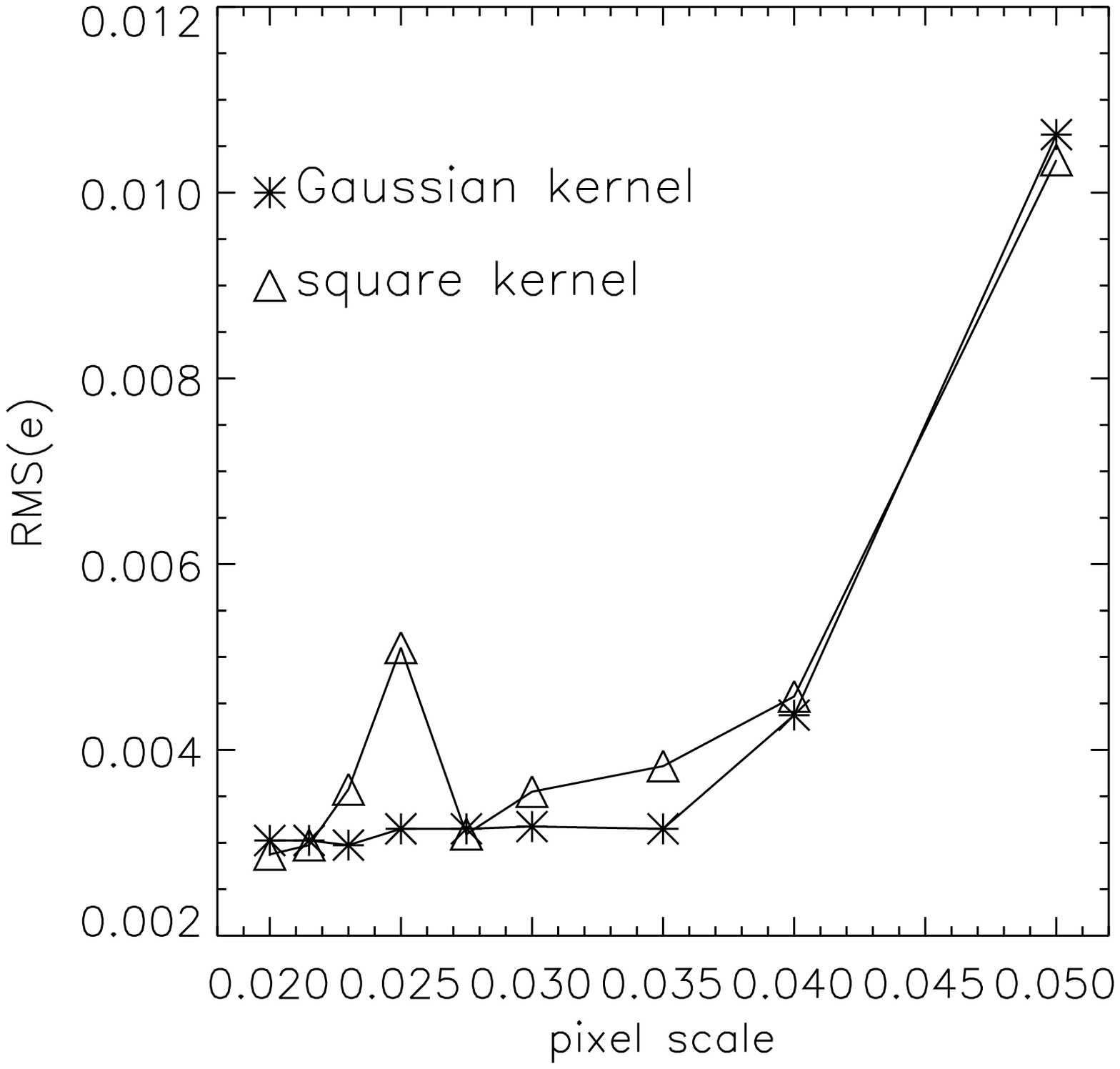}
\caption{RMS ellipticity introduced during the process of removing geometric
distortion and combining dithered images, for a range of \multidrizzle\
parameters. Lower values show more stable behavior of the PSF during this
process. Based on this plot, we recommend a Gaussian kernel and an output pixel
size of 0.03 arcseconds in order to minimize the effect of undersampling on
the PSF and produce images that are optimal for weak lensing science. We note
that the Gaussian kernel introduces significant additional correlated noise
relative to the square kernel, which is not important for weak lensing science
but may not be optimal for other types of science.}
\label{fig:multidrizzle}
\end{figure}

We then ran a series of tests using \multidrizzle\ on the same input image
but with a range of output pixel scales, convolution kernels, and values of
\texttt{pixfrac}. We then measured the scatter in the ellipticity values in the
output images. The smaller that scatter, the more accurately the PSF is
represented. We found the results were not strongly dependent on the choice
of\texttt{pixfrac} and settled on a value of 0.8 for that parameter.  We show in
Figure~\ref{fig:multidrizzle} that PSF stability is improved dramatically by reducing
the output pixel scale from 0.05 arcseconds (the default) to 0.03 arcseconds.
There is only a very slight gain in going to smaller output pixel sizes and the
storage requirements rapidly become problematic. The gain in going to smaller
pixel scales is more stable with a Gaussian kernel for than
with the default square kernel. Therefore, for weak lensing work, we recommend an output pixel scale of
0.03 arcseconds, \texttt{pixfrac}=0.8, and a Gaussian kernel in order to best
preserve the PSF during this stage of image reduction. We note that, despite its
clear advantages for weak lensing studies, the Gaussian kernel does have some
general drawbacks, such as the introduction of more correlated
noise which may not be desirable for other types of science where minimization
of correlated noise is important.

\section{Quantification of the PSF and focus variability} \label{models}

We can measure HST's effective focus at the time of an observation using
$\sim10$ fairly bright stars in the field. We first  created dense grids of
artificial stars across the ACS field of view. By changing the separation of the
primary and secondary mirrors in \tinytim's raytracing model, these models
were made at successive displacements of the focus from nominal, from $-10\mu$m
to $+5\mu$m in $1\mu$m increments. These are reasonable bounds on the maximum
extent of physical variations in HST. The stars are created without geometric
distortion, to avoid the noise that would have been introduced had it been
necessary to carry out geometric transformations on the stellar fields. We then compare the $\sim10$ bright stars
in each COSMOS field to the \tinytim\ PSF grids at each focus value.
Minimizing the difference in ellipticity between the models and the data  finds
 the best fit value of telescope focus for that particular field. Tests on
observations of dense stellar fields that contain many suitable stars show
that this procedure is repeatable with an rms error $\sim1\mu$m, when using ten
different stars repeatedly selected at random.

\begin{figure}
\plotone{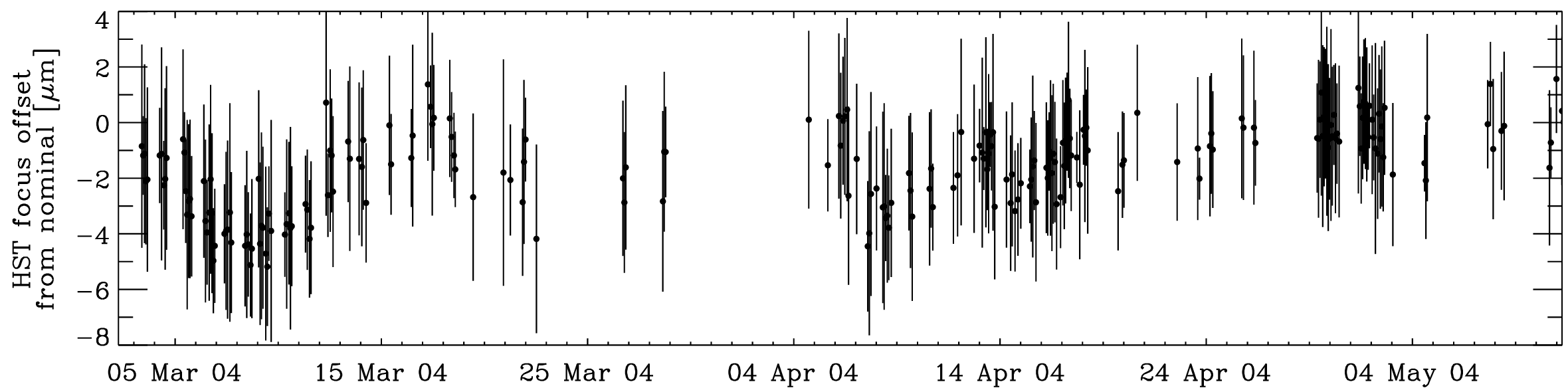} \\
\plotone{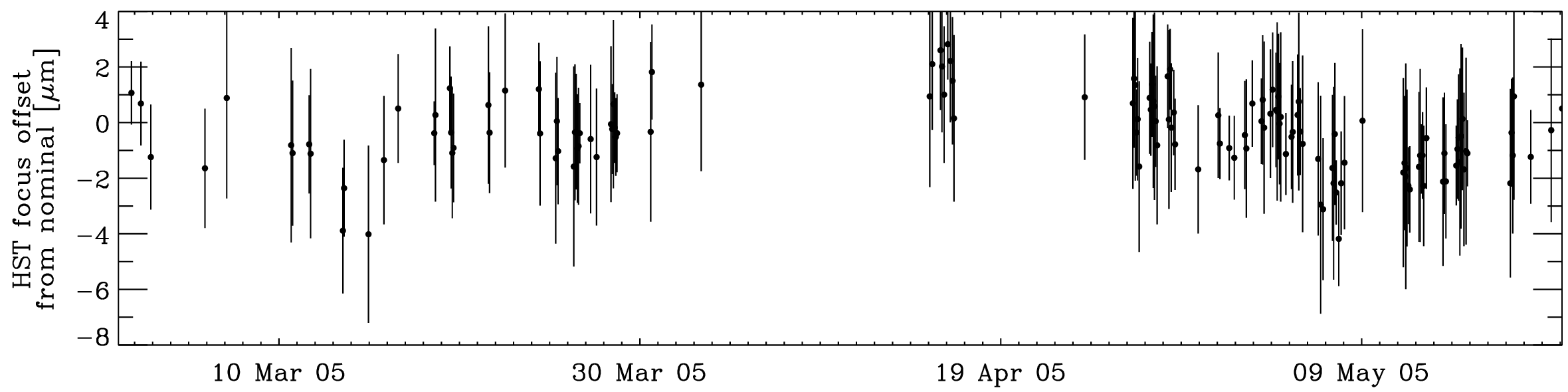}
\caption{Apparent offset of HST from nominal focus during COSMOS observations in cycles 12 and 13. We describe the procedure by which we estimate the telescope's focus in \S\ref{models}}
\label{fig:focus}
\end{figure}

Figure~\ref{fig:focus} shows our estimation of HST's focus in microns away from
nominal during several months in Cycles~12 and 13, using a uniform set of COSMOS
images. HST was not manually refocussed during this time, but the apparent focus
still oscillates. At times, the oscillations seem periodic, but there are also
sharp jumps and more erratic behavior. The random component probably depends in
a complex fashion upon the orientation of HST with respect to the sun and the
Earth during preceding exposures, and we do not believe that it can be easily
predicted in advance.

Note that the uncertainty on the focus value during any single pointing is quite
large; more so than the tests on dense stellar fields would suggest. A major
component of this error is undoubtedly the $\sim3\mu$m thermal fluctuations in
focus that HST experiences during each orbit due to ``breathing''. The COSMOS images are all taken
with a total exposure time of one orbit, and the apparent focus therefore
represents the integral of a gradually changing PSF. We have not been able to
investigate focus changes on short time periods and, given this behavior, it is
even more curious that long term patterns are so clearly present in
Figure~\ref{fig:focus}.

\begin{figure}
\plotone{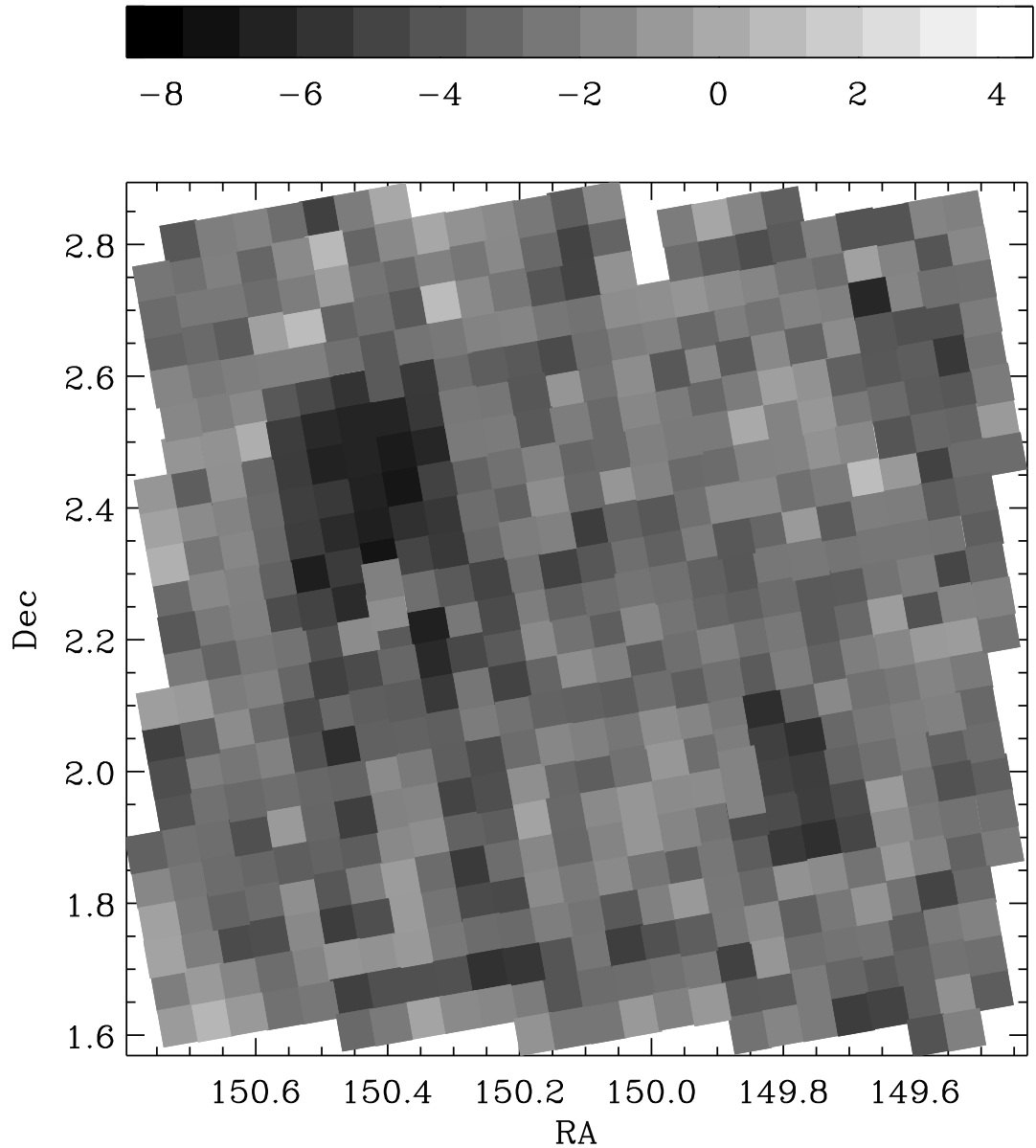}
\caption{Apparent offset of HST from nominal focus in all of the COSMOS fields. COSMOS was taken in chunks,
a few fields at a time.  Therefore, the focus values cluster because fields
taken close to each other in time tend to have similar focus values.  Despite
having only about 10 stars per COSMOS field which are suitable for measuring
PSF, it is apparent from the clustering of focus values in this plot that we can
make a decent estimate of the focus value for individual COSMOS fields using the models and techniques outlined in this paper.}
\label{fig:focuspattern}
\end{figure}

Figure~\ref{fig:focuspattern} shows the focus values determined for all of the
COSMOS fields. The obvious clustering of focus values in that plot is due to the
observing strategy used for COSMOS, in which data was typically taken in chunks of a few fields at a time. Adjacent fields are likely to have been taken at similar times, and therefore tend to have a similar focus values.

\begin{figure}
\plottwo{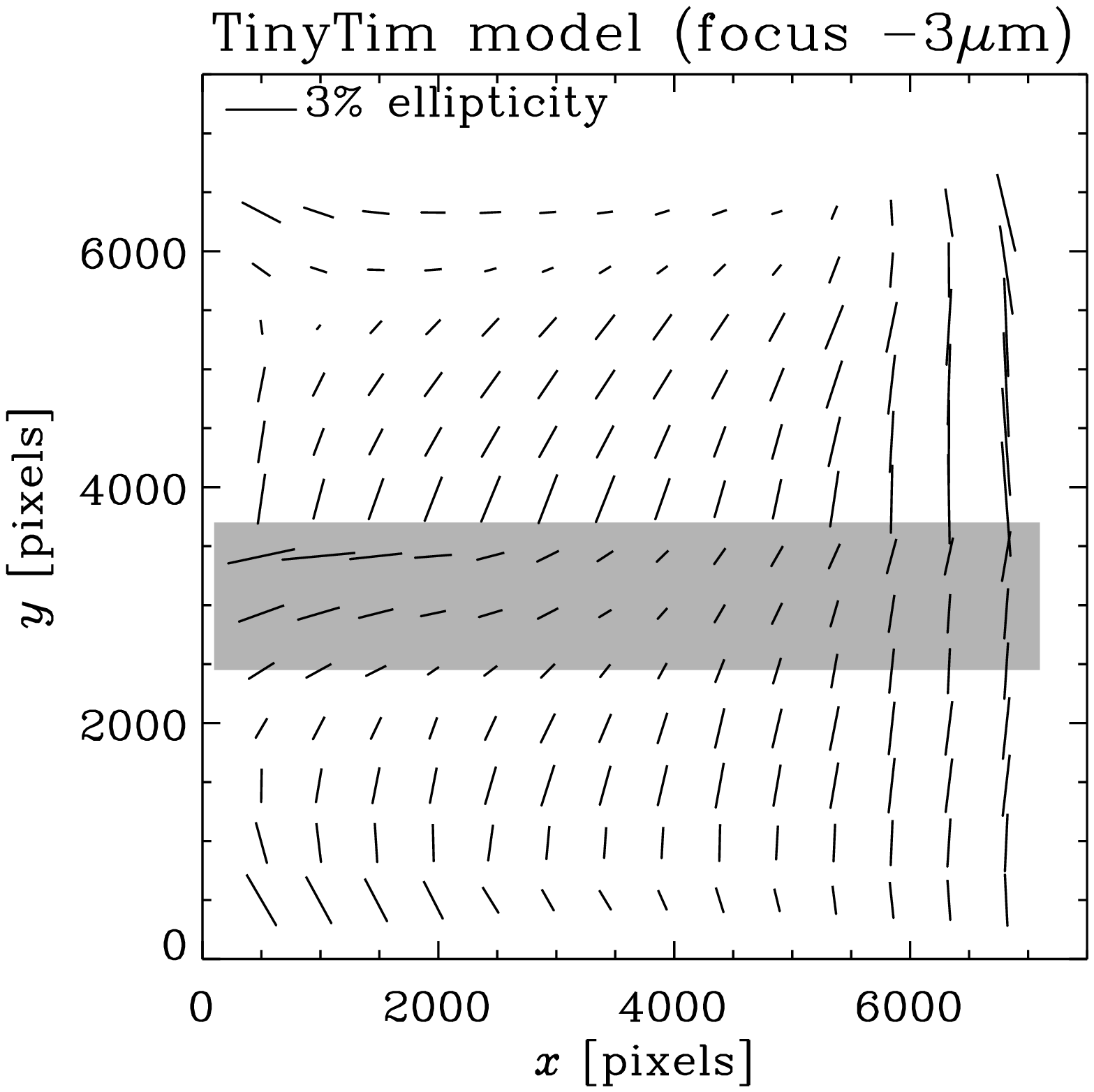}{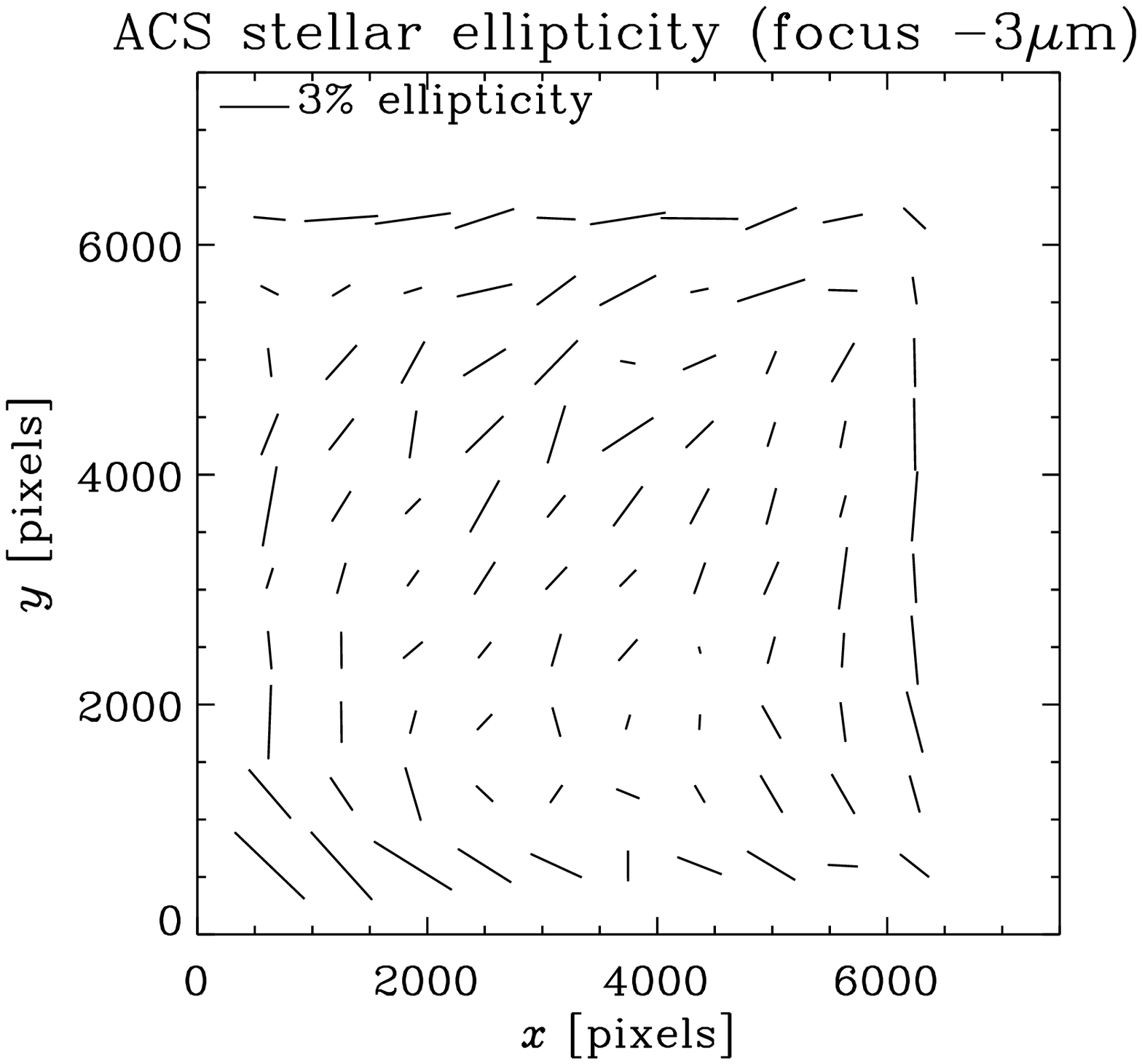}
\caption{
The \tinytim\ PSF models (left panel) for a focus value of $-3\mu$m and observed stars (right panel) in COSMOS fields with a similar apparent focus.  There is good agreement between the data and the models over much of the ACS field. The shaded area near the center of the chip does not show good agreement and this is likely due to the effects of degradation of the Charge Transfer Efficiency (CTE) in the ACS WFC CCDs.}
\label{fig:comparison}
\end{figure}

Figure~\ref{fig:comparison} shows the \tinytim\ models at focus $-3\mu$m and
all the stars from the COSMOS fields with a best-fit focus value of $-3\mu$m.
The COSMOS stars have been averaged in a spatial grid of approximately 600 $\times$ 600 0.03
arcsecond pixels.  There is good agreement between the models and the stars over
most of the field.  The agreement is not as good in the boxed area near the
center of the field.  We believe this is due to a degradation of the CTE of the
ACS CCDs.  This degradation causes trailing of low flux objects in the readout
direction (the $y$ direction).  The effect is most pronounced the further away
the object is from the readout registers at the bottom and top of the field
(Mutchler \& Sirianni 2005; Riess \& Mack 2004).  This causes the objects to be
elongated vertically. Thus, fainter COSMOS stars appear more elongated in the
$y$ direction at the center of the field than the \tinytim\ models, which do
not include the effects of CTE. Note that this does not affect the estimation of
focus positions, because the bright stars matched to our PSF models are less
affected by CTE than the faint sources. We are currently exploring ways to
correct for CTE in all objects, and will publish the results in Rhodes et al.\
(2006).

\section{Conclusions and Future Work}
\label{conclusions}

We have shown that \tinytim\ can produce model PSFs that are very close to
those observed in real data (for example the COSMOS 2-Square Degree Survey).
This required some modifications to the \tinytim\ code, most importantly
adding the ability to mimic the distortion correction and dithering normally
implemented via \multidrizzle, and to produce grids of PSFs across the
entire ACS WFC field of view.  We used \tinytim\ model stars to find the
best values for \multidrizzle\ and found that using a Gaussian kernel,
\texttt{pixfrac }$ =0.8$, and an output pixel scale of 0.03 arcseconds greatly
reduced the ``aliasing'' of point sources introduced during repixellization.

Discrepancies between our models and the COSMOS data can be attributed largely
to a degradation in the ACS CTE since launch. We are currently studying this
problem and will present a complete PSF solution including how to correct for
CTE in Rhodes et al.\ (2006).  We plan to correct science images for CTE on a
pixel-by-pixel basis, like the Bristow code developed for STIS (Bristow et al./ 2002),
moving charge back to where it belongs, rather than including the effects of CTE
in our model PSF (like Rhodes et al. 2004). Thus, the model PSFs
we present here are the ones we plan to use in our weak lensing analysis.

At the time of press, we have thoroughly tested PSF models in only the F814W
filter. However, our IDL routines preserve \tinytim's ability to create
PSFs in other filters, and for sources of any colors. Our routines are therefore
easily adaptable to other data sets.

The whole method is intentionally designed to be as adaptable as possible for
many methods. The desire to know the PSF at any arbitrary position on the sky is
far from unique to weak lensing. But even in lensing, advanced methods like
Shapelets will, in the near future, be able to take advantage of more detailed
information about the PSF shape than it is reasonable to expect from
interpolation between a few stars (Massey \& Refregier 2005; Refregier \& Bacon
2004). This is even more exaggerated when considering higher order shape
parameters, with an intrinsically lower signal to noise. The creation of
noise-free, oversampled stars at any position on an image allows such analysis
in any ACS data.

In the near future, we plan to release our PSF models to the community along
with the wrapper we have written for\tinytim\ which will allow users to
create PSF models in different filters and at user-defined positions in the ACS
field of view. Interested readers are advised to contact the authors for these
resources.

\acknowledgements  We would like to thank Catherine Heymans for sharing her
knowledge of the ACS PSF. We are grateful to John Krist for guiding our poor
lost souls through the underbelly of \tinytim. Great thanks go to Andy
Fruchter and Marco Lombardi for useful discussions about \multidrizzle.
Adam Riess and Marco Sirianni provided expert knowledge about CTE effects. Richard Ellis and Alexandre Refregier
have engaged us in many useful and interesting discussions about the COSMOS field. We
are also pleased to acknowledge the continuing support of Nick Scoville, Patrick
Shopbell and the whole COSMOS team in obtaining and analyzing the COSMOS images
that were used to test our PSF models.

\end{document}